\documentclass{nature}
\usepackage[latin9]{inputenc}
\usepackage{amsmath}
\usepackage{amssymb}
\usepackage{graphicx}
\usepackage{bm}

\makeatletter
\@ifundefined{date}{}{\date{}}

\usepackage{epstopdf}
\usepackage{color}
\usepackage{ulem}
\usepackage{endnotes}
\usepackage{eucal}
\usepackage[dvipsnames]{xcolor}

\usepackage{multirow}

\usepackage{float}

\usepackage{times}

\newenvironment{sciabstract}{%
\begin{quote} \bf}{\end{quote}}

\newcommand{\ket}[1]{|#1\rangle}

\newcounter{lastnote}

\title{Atomic-scale structure analysis of a molecule \\at a (6-nanometer)$^3$ ice crystal}

\author
{Xi Kong,$^{1,2,3,\dag}$
Fazhan Shi,$^{1,2,3,\dag}$
Zhiping Yang,$^{1,\dag}$
Pengfei Wang,$^{1,3,}$
Nicole Raatz,$^{4}$\\
Jan Meijer,$^{4}$
Jiangfeng Du$^{1,2,3,\ast}$\\
\\
\normalsize{$^{1}$CAS Key Laboratory of Microscale Magnetic Resonance and Department of Modern Physics,}\\
\normalsize{University of Science and Technology of China (USTC),}
\normalsize{Hefei, 230026, China}\\
\normalsize{$^{2}$Hefei National Laboratory for Physical Sciences at the Microscale, USTC}\\
\normalsize{$^{3}$Synergetic Innovation Center of Quantum Information and Quantum Physics, USTC}\\
\normalsize{$^{4}$Felix-Bloch institute for solid state physics University Leipzig, }\\
\normalsize{Linn\'{e}str. 5, D-04103 Leipzig, Germany}
\\
\normalsize{$^\dag$ These authors contributed equally to this work.}\\
\normalsize{$^\ast$ Corresponding author. E-mail: djf@ustc.edu.cn.}
}

\makeatother

\begin{document}

\baselineskip24pt


\maketitle


\begin{sciabstract}

Water is the most important solvent in nature. It is a crucial issue to study interactions among water molecules. Nuclear magnetic resonance (NMR) spectroscopy is one of the most powerful tools to detect magnetic interactions for the structure analysis of a molecule with broad applications \cite{wuthrich_way_2001,gossuin_physics_2010}.
 But conventional NMR spectroscopy requires macroscopic sample quantities with hampers in investigating nanoscale structures \cite{glover_limits_2002}.
Through quantum control of a single spin quantum sensor, magnetic resonance spectroscopy of nanoscale organic molecules\cite{staudacher_nuclear_2013,mamin_nanoscale_2013} and single molecules\cite{shi_single-protein_2015,lovchinsky_nuclear_2016} has been achieved.
However, the measurement of the dipolar interaction of nuclear spins within a molecule at nanoscale and the analysis of its structure remain a big challenge.
Here we succeed in detecting the NMR spectrum from an ice crystal with (6-nanometer)$^3$ detection volume.
More importantly, the magnetic dipolar coupling between two proton nuclear spins of a water molecule was recorded. The resolved intra-molecule magnetic dipolar interactions are about  15 kHz and 33 kHz with spectral resolution at a few kHz.
Analysis of the interaction-resolved NMR spectroscopy provides a spatial view of nanoscale ice crystal, from which the orientation of a water-molecule bond is derived and further the length of the bond can be got.
This work enables NMR spectroscopy applications in single molecule structure analysis, provides a further tool for nanocrystalline and confined water research \cite{algara-siller_square_2015,guo_nuclear_2016}.

\end{sciabstract}

Water is the most important solvent in nature. The interaction among water molecules is a very crucial issue, for example it gives rise to the  ``hydrophobic force" which is responsible for membrane formation and contributes to protein structure. Yet, information on how water structure is on nanoscale is scarce. On the one hand it makes hard to observe in X-ray\cite{algara-siller_square_2015} or electron microscopy\cite{adrian_cryo-electron_1984} that water comprises light elements. On the other hand bulk methods like dielectric spectroscopy\cite{marcus_ion_2006} does not allow to access local information which is essential when water is interacting with solutes. Here we use nanoscale NMR to provide an unprecedented insight into water structure formation. Because conventional NMR spectroscopy requires macroscopic sample quantities\cite{glover_limits_2002}, extending NMR spectroscopy to structure analysis of molecule at nanoscale is a long standing goal.
Recently, a single quantum spin sensor, the nitrogen-vacancy (NV) defect in diamond, has been developed to realize nanoscale magnetic resonance spectroscopy\cite{balasubramanian_nanoscale_2008,maze_nanoscale_2008}.
During the last several years, magnetic resonance spectroscopy of nanoscale organic molecules\cite{staudacher_nuclear_2013,mamin_nanoscale_2013} and single molecules\cite{shi_single-protein_2015,lovchinsky_nuclear_2016} has been achieved.
However, the measurement of the dipole-dipole interaction of nuclear spins within a molecule at nanoscale and the analysis of its structure remain ellusive.
In this work, we report the NMR spectrum from an ice crystal at (6-nanometer)$^3$ detection volume. More importantly, the magnetic dipolar coupling between two proton nuclear spins of a water molecule was recorded.
The analysis of the interaction-resolved NMR spectra provides a spatial view of the ice crystal and the structure of water molecules inside.

\begin{figure}
\centering
\includegraphics[width=1.0\columnwidth]{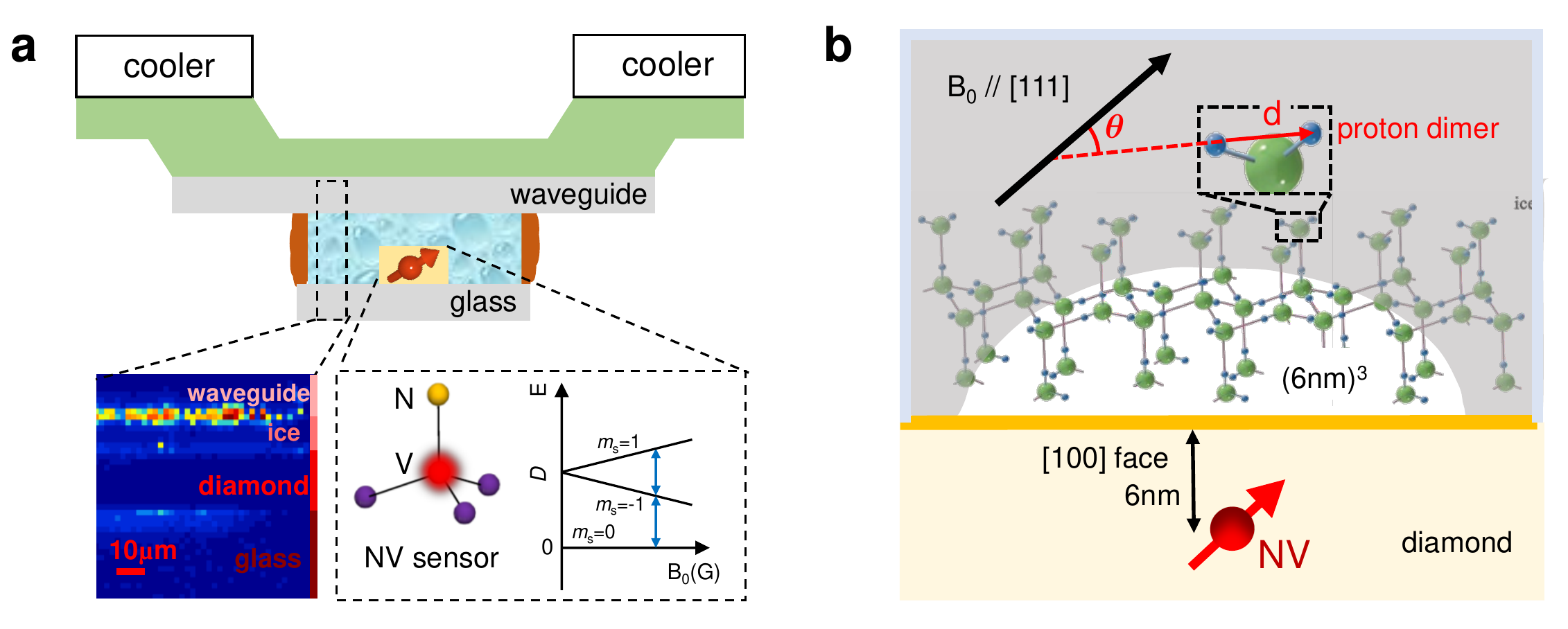}
\linespread{1.5}
    \caption{\textbf{Schematic of the experimental setup.}
    \textbf{a,} Proton nuclear spins in nano-scale ice are coupled to an NV sensor which is readout by confocal optical system. The ice is made by freezing a small volume of water which is filled in between the gap of diamond and coplanar waveguide (thickness is about 10$\mu{}$m). A semiconductor cooler is mounted above the coplanar waveguide with a copper plate to make a good heat contact with the sample. The sample temperature is controlled down to 266K(-7$^{\circ}$C, see Supplementary Information for details). Left inset shows the profile configuration of our setup. Middle inset shows the structure of the magnetic sensor, nitrogen-vacancy (NV) center, which is consist of a substitute nitrogen atom and an adjacent vacancy. Right inset shows the energy level of the spin ground state of NV center under static magnetic field B$_0$. \textbf{b,} A detail schematic of the detection configuration. The shallow NV center as magnetic sensor in diamond (6 nm depth) couples to nearby protons in the ice layer. The sensing volume of the NV sensor is about (6nm)$^3$ of proton nuclear spins (contributing more than 90\% of the signal) in ice above the sensor. The ice crystal is hexagonal crystalline structure denoted as ice I$_h$. The external magnetic field is $54^{\circ}$ tilted from the surface to align with the main axis of NV sensor. The angle between magnetic field B$_0$ and the proton dimer bond in a water molecule is $\theta$, while the distance between two protons is $d$.
 }\label{fig:1}
\end{figure}

The NV center is a highly sensitive atomic-scale magnetic sensor \cite{Hollenberg2013review, Degen2016arxiv}. It consists of a nitrogen impurity and a neighbor vacancy in diamond (Fig. 1a).
A spin triplet ground state of an NV center can be initialized and readout by 532 nm illumination.
Such a physical system can be used for detecting magnetic fields.
It performs high sensitivity and high resolution spin spectroscopy of targets both in diamond\cite{zhao_sensing_2012,shi_sensing_2014} and near surface\cite{staudacher_nuclear_2013,mamin_nanoscale_2013,shi_sensing_2014,lovchinsky_nuclear_2016,muller_nuclear_2014,sushkov_magnetic_2014,staudacher_probing_2015}.
In our experiments, NV centers are implanted by 5 keV N$^{+}$ ions into a diamond of 50 \textit{$\mu{}$}m thickness.
The depth of shallow NV centers are identified by NMR based methods to measure the distance to  protons to be 5-7 nm  (Tabel S1 in Supplementary Information). The experiment setup is shown schematically in Fig. \ref{fig:1}a.
The magnetic sensor in diamond is mounted between a coplanar waveguide and a glass plate.
The water is filled in the gap between the diamond and the coplanar waveguide, while paraffin wax is dropped around the gaps to prevent water from evaporation. The whole sample together with magnetic sensor and waveguide is connected to two semiconductor coolers in nitrogen gas atmosphere.
The water sample is frozen to solid and its structure is a hexagonal crystal of \textit{I}$_{h}$, as shown in Fig \ref{fig:1}b.
The sensing volume of the protons in ice is around $(6\textrm{nm})^3$ (Fig. \ref{fig:1}b).

The sensor-sample system Hamiltonian is $H=H_{\textrm{NV}}+H_{\textrm{hf}}+H_{\textrm{nuc}}$. The NV sensor spin Hamiltonian is $H_{\textrm{NV}}=D S_z^2+\gamma_e {\textbf{B}}\cdot {\textbf{S}} $, where $D$ denotes the zero field splitting and $\gamma_e=2.8$MHz/G is the electron spin gyromagnetic ratio.
The NV sensor couples to the proton nuclear spins through the hyperfine Hamiltonian, $H_{\textrm{hf}}=S^z\sum_{m=1}^N\left({A^{zz}}_mI_m^z+{A^{zx}}_mI_m^x\right)$, where $A^{\alpha\beta}_m$ is the hyperfine tensor, $S^z$ and $I_m^{\alpha}$ is NV spin and proton nuclear spin, respectively.
The proton nuclear spins Hamiltonian is $H_{\textrm{nuc}}=\omega_LI_m^z+ \sum_{m=1}^N\sum_{n=1}^{m-1}{{\vec{I}}_m\cdot{\overleftrightarrow{\mathfrak{D}}{}}}_{m,n}\cdot{\vec{I}}_n $, where \textit{$\omega{}$}$_{L }= \gamma_H B_0$ is the Larmor frequency of nuclear spins, ${\overleftrightarrow{\mathfrak{D}}}_{m,n}$is dipolar interaction between nuclear spins $m$ and $n$.

\begin{figure}
\centering
\includegraphics[width=1\columnwidth]{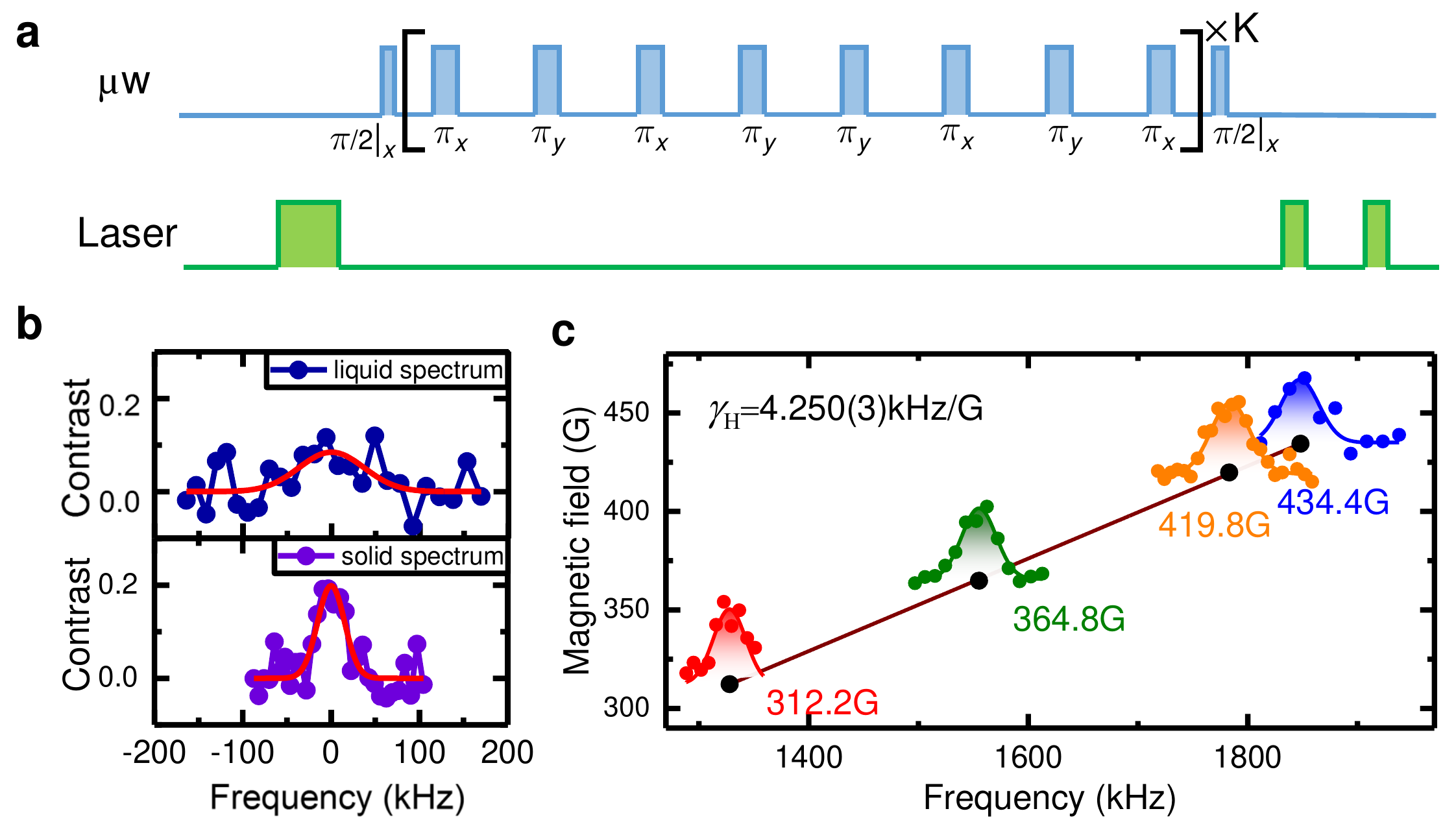}
\linespread{1.5}
\caption{\textbf{Nuclear Magnetic Resonance spectrum of protons in water molecules.}
\textbf{a,} Schematic protocols of NV sensor to detect water molecules.
The NV center is firstly initialized into $|m_s=0\rangle$ state by optically pumping, followed by a resonant microwave $\pi{}$/2 pulse transforming the NV spin state in a superposition state $(|m_s=0\rangle+|m_s=-1\rangle)/\sqrt{2}$. Then we drive the transition $\vert{}\left.m_s=0\right\rangle{}\leftrightarrow\vert{}\left.m_s=-1\right\rangle{}$ of  the NV spin with a periodic train of microwave \textit{$\pi{}$} pulses.
Afterwards another $\pi/2$ pulse transforms the NV sensor spin and then readout population in $\ket{m_s=0}$ by green laser pumping.
\textbf{b,} Liquid NMR spectrum and solid state NMR spectrum of proton nuclear spin is plot in purple curve. The line-width of liquid NMR is 87$\pm$36kHz and solid NMR is 36$\pm$5kHz. \textbf{c,} NMR spectrum of ice protons as a
function of applied magnetic field, using the XY8-12 dynamical decoupling sequence (96 $
\pi$ pulses). The resonant
frequency is taken from a fit at magnetic fields of 312.2G, 364.8G, 419.8G, 434.4G. The
detected signal has a slope of 4.250(3) kHz/G which matches with the gyromagnetic
ratio of proton (purple solid line).
 }\label{fig:2}
\end{figure}

To detect the NMR signal of protons in water, a ``lock-in-detection" method is used.
A periodic dynamical decoupling pulse sequence, XY8-K is used to control the NV center.
When the sensor is driven in synchrony with nuclear evolution ( $\pi$ pulse intervals $\tau$ are adjusted to \textit{$\tau{}$}=1/(2\textit{$\omega{}$}$_{L}$) in Fig. \ref{fig:2}a), the effective evolution of nuclear spin is given by $A^{zx}_m I^x_m/\pi$ \cite{muller_nuclear_2014}.
An a.c. magnetic signal from nuclear spins causes the decoherence of sensor spin state, which is then readout optically after another $\pi{}$/2 pulse.
By sweeping $\tau$ and converting it to frequency domain (Fig. S6),  NMR spectra are observed ( Fig. \ref{fig:2}b ).
The NMR spectrum of water at room temperature is shown in the upper panel in Fig. \ref{fig:2}b.
Surprisingly, the full-width at half maxium (FWHM) of the  liquid spectrum is much broader than the liquid spectrum from a conventional 400MHz-NMR spectrometer (Fig. S5). Even the liquid spectrum is broader than the solid spectrum by a same NV sensor ( lower panel in Fig. \ref{fig:2}b ), which is opposite to conventional NMR spectra.
In fact, it is a special phenomenon in nano-NMR spectroscopy due to the diffusion of sample at nanoscale sensing volume.
The spectral broadening of liquid water comes from the fast diffusion of water molecules through the detection volume of the NV sensor \cite{staudacher_probing_2015}.
To eliminate the diffusion and preserve the dipolar interaction, the water is frozen to solid.
The spectral FWHM of frozen water is about 36 kHz, which mainly results from the dipolar interactions and the detection sequences.
It is comparable with the solid spectrum from a conventional spectrometer.
NMR spectra at various fields are observed (Fig. \ref{fig:2}c).
The fitted resonant frequencies are proportional to the external magnetic fields. The observed gyromagnetic ratio 4.250(3) kHz/G matches well with proton nuclear spins.
All spectra are verified with correlation spectroscopy (Fig. S8 and Ref.\cite{loretz_spurious_2015}).

\begin{figure}
\centering
\includegraphics[width=0.95\columnwidth]{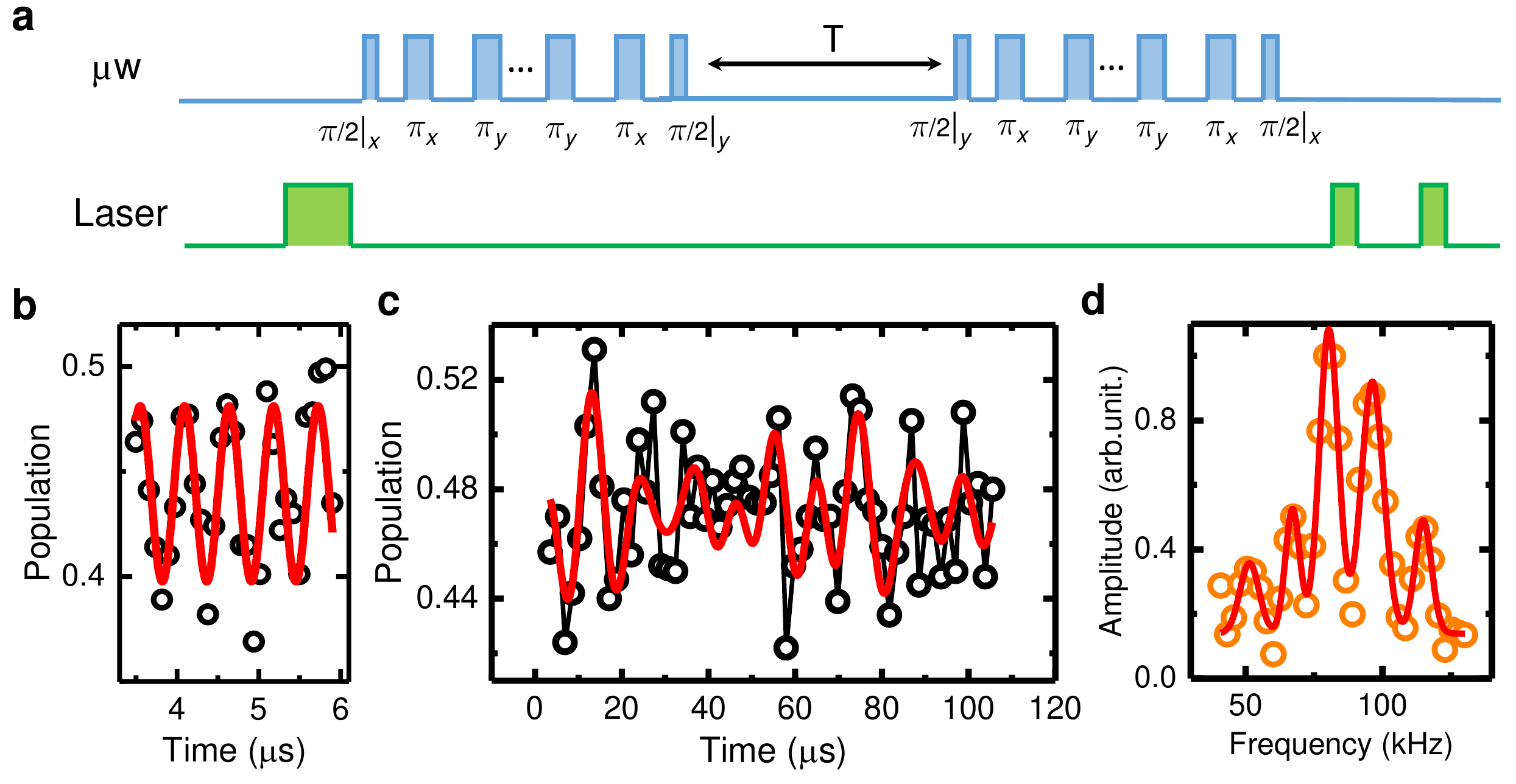}
\linespread{1.5}
\caption{\textbf{Correlation spectrum of protons in ice and its Fourier transformation spectrum.}
\textbf{a,} Schematic of the control protocol to resolve the coupling of proton spins. After the first pumping laser, a $\pi/2$ in x axis rotate the sensor spin to transverse plane. Subsequently a periodical train of 8$K$ $\pi$ pulses flip the sensor spin with time intervals between $\pi$ pulses $\tau=1/2\omega_L$. Following another $\pi/2$ pulse along y axis correlate the NV sensor spin with proton nuclear spins in ice. Another $XY8-K$ sequence is applied after $T$ to observe the free evolution of nuclear spins between $T$ time.
\textbf{b,} The sequence in (\textbf{a}) is applied to an NV sensor covered by ice. The proton evolves with a Larmor frequency $\gamma_HB_0$=1847 kHz at an external magnetic field B=434 G. From the correlation spectroscopy we find the oscillation period $\tau_L=544\pm 8$ns match with Larmor frequency of proton.
\textbf{c,} We carry out an under-sampling protocol to record the free evolution envelope of proton nuclear spins. A modulation is observed as a combination of five frequency components $\omega_{1,2,3,4,5}$= 51.1kHz, 67.1kHz, 79.0kHz, 97.1kHz, 114.8kHz. The results fit well with a sum of five sinus functions as shown.
\textbf{d,} FFT of correlation spectroscopy is shown. The five peaks are fitted with five frequencies $\omega_{1,2,3,4,5}$= 51.5kHz, 67.2kHz, 80.4kHz, 96.3kHz, 114.9kHz which meet well with the fitting of time domain curve.
}\label{fig:3}
\end{figure}

The dipolar coupling is usually less than a hundred kHz.
To detect this kind of weak coupling, a high-resolution strategy, correlation spectroscopy\cite{staudacher_probing_2015,boss_one-_2016,laraoui_high-resolution_2013,kong_towards_2015}, is then carried out to resolve the nuclear spin interactions.
With the periodic dynamical decoupling method, the coherence time $T_2$ of NV centers with depth of a few nanometers is at the order of a few tens of microseconds.
The correlation spectroscopy extends the frequency resolution from 1/\textit{T}$_{2}$ to 1/\textit{T}$_{1}$, which is about a few kHz for most of shallow NV centers, sufficient to resolve the intra-molecule nuclear spin dipolar interaction. The pulse sequence is described in Fig. \ref{fig:3}a.
The protocol consists of two dynamical decoupling sequences with the free evolution time $T$ in between \cite{boss_one-_2016}.
Both of the dynamical decoupling sequences are applied to correlate the NV sensor with proton nuclear spins. During the free evolution time $T$, the protons evolve freely under the external magnetic field and the mutual proton-proton dipolar interactions.

The method  allows us to measure the free evolution of the transverse proton spin  for an extented time interval $T$ as shown in Fig. \ref{fig:3}b. The frequency is 1838 $\pm$ 27 kHz which matches with Larmor frequency $\gamma_H \text{B}_0$ = 1847 kHz of proton nuclear spins under the external field B$_0$.
To resolve the weak couplings, a correlation protocol with $T$ = 110$\mu$s is taken to detect the NMR spectrum with a line-width on order of $\sim$ 5 kHz.
Without losing any 
spectral information, an under-sampling protocol is carried out to record the evolution envelope of proton nuclear spins.
A modulation is observed on the correlation spectrum (Fig. \ref{fig:3}c) and its fast Fourier transform (FFT) shows the frequency components more straightforward (Fig. \ref{fig:3}d).

\begin{figure}
\centering
\includegraphics[width=1\columnwidth]{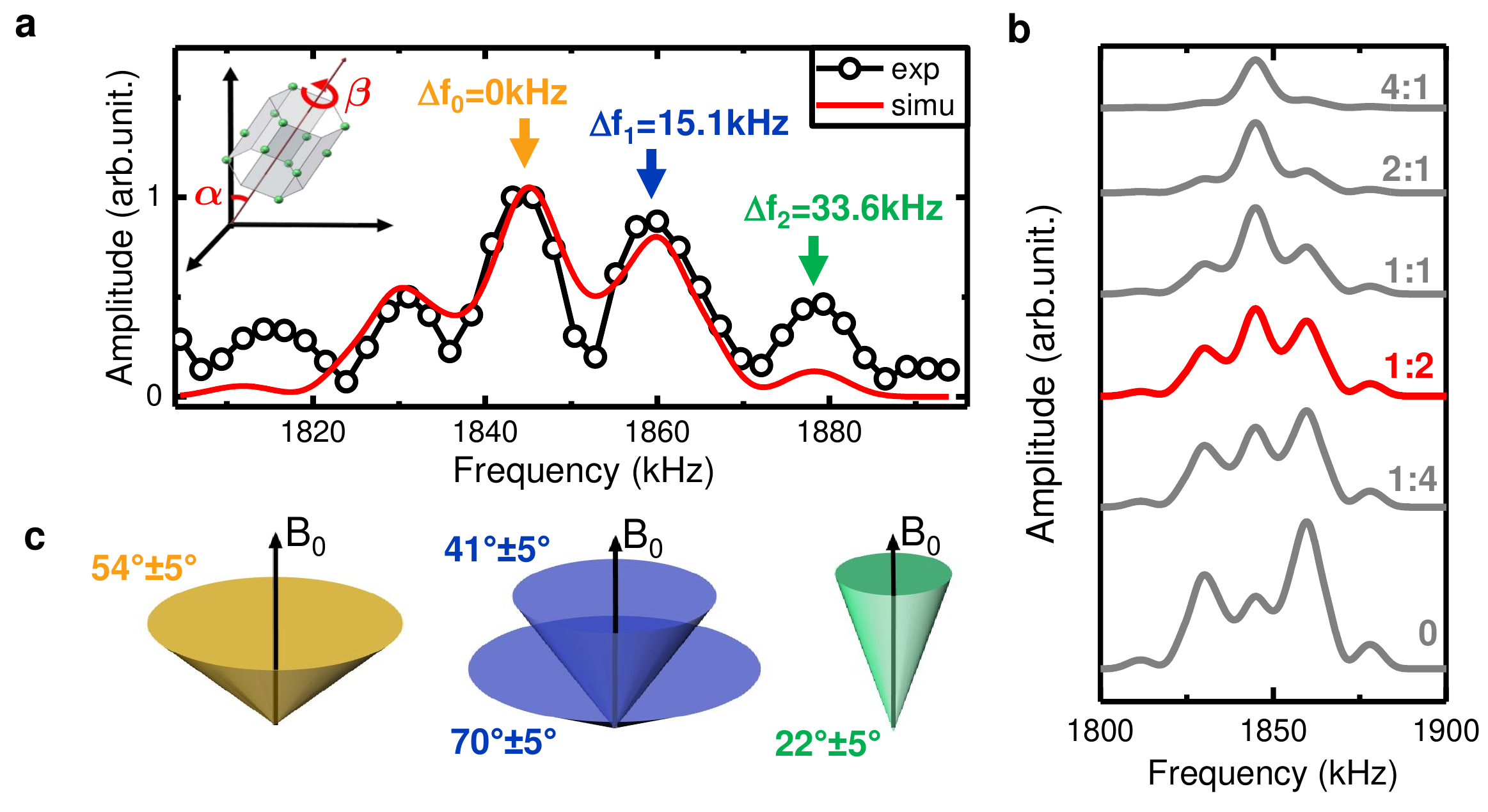}
\linespread{1.5}
\caption{\textbf{Structure analysis of a molecule in nanoscale ice crystal by simulations of the NMR spectra.} \textbf{a,} The original spectrum from Fig.\ref{fig:3}\textbf{d} is shown, the frequency axis is reconstructed as described in main text. The orange, blue and green arrows show the spectra splittings with $\Delta f_{0,1,2}= 0\textrm{kHz}, 15.1\textrm{kHz}, 33.6\textrm{kHz}$, which correspond to different proton dimer bond orientations. The orange arrow label the mixture peaks of HDO molecules spectra and H$_2$O molecules with $\theta=54^{\circ}$ tilted from external magnetic field.
Simulated spectrum of proton-NMR fits well with experimental curve with optimal crystal configuration $\theta=65^{\circ}\pm 2^{\circ}$, $\phi=79^{\circ}\pm 2^{\circ}$.
\textbf{b,} We simulate the spectra with various ratio of semi-heavy water molecules HDO to light water molecules H$_2$O range from 0 to 4:1. The gyromagnetic ratio of a deuterium nuclear spin is 0.6536 kHz/G.
\textbf{c,} The three spectra splittings are corresponding to different proton dimer orientations distributing on the color cones (Table\ref{theta}).  The $\Delta f_{0}$ peak corrspond to the orange cone with $\theta=54^{\circ}\pm 5^{\circ}$ tilted with external magnetic field. The $\Delta f_{1}$ peak is a mixture of $\theta=41^{\circ}\pm 5^{\circ}$ and $\theta=70^{\circ}\pm 5^{\circ}$ proton dimer orientations as shown in blue cones. The third $\Delta f_{2}$ peak corresponds to $\theta=22^{\circ}\pm 5^{\circ}$ green cone.
}\label{fig:4}
\end{figure}

The original spectrum, shown in Fig. \ref{fig:4}a, is reconstructed from the spectrum in Fig. \ref{fig:3}d according to the Nyquist-Shannon sampling theorem\cite{oppenheim_digital_1975} (see Supplementary Information for details). 
The central peak, marked by yellow arrow with $\Delta f_0 = 0$ kHz, comes from both of zero coupling protons in H$_2$O and the unpaired protons in HDO.
It is further known that the other four peaks correspond to two frequency splittings $\Delta f_{1,2}$= 15.1 kHz, 33.6 kHz, caused by magnetic dipolar interactions of proton nuclear spins.
Through the analysis of the spectrum, we can derive the orientation of an ice nanocrystal, the directions of proton dimers and even the distance $d$ between two protons in a molecule.
The distance $d$ can be extracted by measuring the NMR spectra under different $\theta$.

The homonuclear magnetic dipolar interaction is formulated by
$H_{\textrm{D}}= \delta\left(1-3\cos^2\theta{}\right)(3I_1^z I_2^z - \vec{I}_1\cdot\vec{I}_2)/2$, where the dipolar coupling parameter \textit{$\delta = \frac{\mu_0}{4\pi}\frac{\gamma_H^2\hbar}{d^3}$} and $\theta$ is the angle of proton dimer orientation vectors with external magnetic field B$_0$.
The dipolar splitting frequency relative to Larmor frequency of protons is $\Delta f =\frac{3}{4} \delta \left(1-3\cos^2\theta{}\right)$.
The parameter $\delta{} \propto 1/d^3$ leads to the inter-molecule interaction much smaller than intra-molecule interaction in most cases.
The inter-molecular dipolar interaction in our experiment is further reduced by dilution of proton nuclear spins (1:1 mixture of light and heavy water).
The dipolar coupling parameter $\delta{}$ is approximately 30.5 kHz for a distance $d$ = 1.58{\AA} between protons of a molecule.
The average proton nuclear spin interaction strength decreases from 4 kHz to 2 kHz, which is smaller than the resolution in the experiment.
In the following analysis of the spectrum, only intra-molecule interaction is considered.

We assume that the ice on the diamond surface is a single crystal as our sensor only detects a sample with a few cubic nanometer volume.
Under our experimental conditions, the ice single crystal has $I_h$ symmetry with azimuth angle ($\alpha,\beta$) (inset in Fig. \ref{fig:4}a).
In each \textit{I}$_{h}$ ice crystal cell, there are water molecules with different orientations. In the ice, there are 12 different proton dimer orientations $\theta_i$ in total and all of them depend only on the crystal azimuth angle ($\alpha,\beta$).
The dipolar interaction of a proton dimer bond depends on the angle $\theta_i(\alpha,\beta)$ between the dimer orientation with the external magnetic field, where $i=1, \ldots ,12$.
The spectrum splitting, $\frac{3}{4}\delta{}\left(1-3\cos^2\theta \right)$, is determined by each of the proton dimer angle $\theta_i$ and distance $d$ in Fig. \ref{fig:1}b.

The spectra are calculated by $\sum_{i=1}^{N}\frac{3}{4}\delta{}\left(1-3\cos^2\theta_i\right)$. Matching the calculated spectra with experiments finds the crystal azimuth angle $\alpha,\beta$ and thus determines the proton dimer bond angles $\theta_i$.
The spectra are dominated by proton nuclear spins in both of light water and semi-heavy water molecules as the signal from deuterium is mitigated due to their nearly one order smaller magnetic moment.
The ratio D$_2$O:H$_2$O is originally 1:1, which may slowly decreased due to the absorb and exchange between the liquid water and the water molecules in air. We simulate the spectra with various ratios of semi-heavy water molecules HDO to light water molecules H$_2$O step by step. The simulation curve matches with the experimental spectra well when the molecules ratio HDO:H$_2$O is 1:2 (Fig. 4b).


The crystal orientation ($\alpha,\beta$) = (65$^\circ$, 79$^\circ$) as an optimal azimuth angle is firstly derived (see Supplementary Information for details). 
Thus 12 different proton dimer orientations $\theta_i(\alpha,\beta)$ are resolved and listed in Table \ref{theta}.
Three peaks in the spectrum, $\Delta f_{0,1,2}$, indicated by yellow, blue and red arrows in turn (Fig. \ref{fig:4}a), are contributed by proton nuclear spin dimers with directions $\theta _{1,2}$, $\theta _{3\sim10}$, and $\theta _{11,12}$, respectively.
The orientations are shown as circular conical surface relative to B$_0$ in Fig. \ref{fig:4}c.

In the above analysis, we assume the proton dimer bond length to be 1.58{\AA}. In principle, the distance $d$ can be extracted by measuring the dipolar splitting under different $\theta$.
 The dimer bond length resolution is about 0.1{\AA}, which is estimated from a broadened spectrum with FWHM $\approx$ 6 kHz.

\begin{table}[H] 
\scriptsize
\centering
\caption{\label{theta} the orientations of twelve proton nuclear spin dimers.}
\begin{tabular}{|c||c|c||c|c|c|c|c|c|c|c||c|c|}\hline
\multirow{2}{*}{Coupling / kHz}
&\multicolumn{2}{c||}{$\Delta f_0$} & \multicolumn{8}{c||}{$\Delta f_1$} & \multicolumn{2}{c|}{$\Delta f_2$}\\
&\multicolumn{2}{c||}{0.0$\pm$4.1} & \multicolumn{8}{c||}{15.1$\pm$4.8} & \multicolumn{2}{c|}{33.6$\pm$3.6}\\ \hline
$\theta_{i:1\sim12} $ / $^{\circ}$ &59.5&55.4&65.1&65.1&70.7&70.7&80&81.6&41.3&41.3&22&26\\\hline
$\frac{3}{4} \delta \left(1-3\cos^2\theta_i\right)$ / kHz&5.2&0.8&10.7&10.7&15.3&15.3&20.6&21.3&15.8&15.8&36.1&32.3\\\hline
\end{tabular}
\end{table}

In conclusion, we firstly observe the liquid and solid NMR spectra and firstly resolve the magnetic dipolar coupling of two protons within a water molecule. Through the analysis of the magnetic dipolar coupling spectrum,  we resolved the orientation of an ice nanocrystal with ($\alpha,\beta$) = (65$^\circ$, 79$^\circ$) and the directions of proton dimers $\theta_{i:1\sim12}$ which listed in Table\ref{theta}. We further can get the distance $d$ between two protons in a molecule using the same spectral analysis method. Equipped by new technology with high sensitivity\cite{lovchinsky_nuclear_2016} and angle-adjustable magnetic field\cite{lovchinsky_magnetic_2017}, there'll be more constraints for the crystal orientations. Thus we can uniquely determine the dimer bond length and the angle for both by our method.

Detection of magnetic interactions is an essential part of the nano-NMR spectroscopy.
This work shows that NMR spectroscopy yields the structure analysis of ice crystal and a water molecule at nanoscale by an NV sensor.
This work provides a further tool for single molecule structure analysis.
Together with the previous work on nano-NMR spectroscopy \cite{staudacher_nuclear_2013, mamin_nanoscale_2013}, spectrally resolved dipolar interaction in small sample volume can yield valuable structural information as opposed to bulk NMR, where such interactions typically hamper structure analysis.
Combined with detection of chemical shift, $J$-coupling and widely used nuclear spin labeled methods in conventional NMR, the structure of a single complex bio-molecule could be fully resolved by NV-based nano-NMR spectroscopy.
More importantly, the detection of nanoscale water is a highly challenging scientific field in itself.
As nanoscale proton configurations are especially difficult to observe under ambient conditions for both of STM or X-ray, the NV-based sensing provides a unique way to take a close snapshot of nanoscale configuration.
For these reasons, the NV-based quantum sensing has the potential to shed light on a long standing challenge in structural biology: the role of water layers on protein structures.
Water is the most important solvent in nature and water layers on surfaces or around molecules determine protein folding or the energetics of cell membranes.
For example, the structure of water bound to surfaces such as proteins is of outstanding importance for their function but is only accessible by molecular dynamics simulation. Here we provide a method which enables to detect such structure.
Besides, this work opens a way in the research of nanoscale confined water with a few atomic layers by magnetic resonance, which may reveal novel physical phenomena and gain fundamental knowledge of water related science
\cite{brown_how_2001,chandler_interfaces_2005,verdaguer_molecular_2006,algara-siller_square_2015,guo_nuclear_2016}.

%

\begin{methods}
\section*{cooling system}
The whole sample area is cooled by two semiconductor coolers. The coolers are pasted on a cooper heat sink with silicone grease. The temperature of the cooper sink is measured to be -20$^{\circ}$C when a 17$^{\circ}$C cooling water is carried out on the heat side of the semiconductor coolers. All the setup is mounted in a nitrogen gas atmosphere to prevent frost. The cooper heat sink is glued on the coplanar waveguide with silicone grease to cooling down the water stored between the waveguide and NV sensor in diamond. The lowest temperature measured by NV sensor is -7$^{\circ}$C.

\section*{calculation and fitting}
To simulate the spectrum, we consider the proton nuclear spins to be interacted with adjacent proton or adjacent deuterium nuclear spins. The dominated coupling of  adjacent proton-proton dipolar interaction is $\Delta f(\alpha,\beta)$. The coupling splitting caused by proton-deuterium dipolar interaction is $\Delta f^{\textrm{D}}(\alpha,\beta)$. There're portion $p$ deuterium-proton water molecule, such that there would be $p/(1-p)$ deuterium peak compared with only proton-proton peak. Considering the peak broadening $\Delta$, frequency shift $f_{\textrm{shift}}$ and linewidth $\sigma$ of the filter function, we have the fitting function,
\begin{equation}
\begin{aligned}
S(\alpha,\beta,f_{\textrm{shift}}) &= \bigg( \frac{2p}{1-p}\sum_{i=1}^{12}\frac{1}{3}
\Big(e^{-(f-\Delta f^{\textrm{D}}(\alpha,\beta))^2/\Delta^2}+e^{-f^2/\Delta^2} +e^{-(f+\Delta f^{\textrm{D}}(\alpha,\beta))^2/\Delta^2}\Big)
\\
 &+\sum_{i=1}^{12}\left( e^{-(f-\Delta f(\alpha,\beta))^2/\Delta^2} +e^{-(f+\Delta f(\alpha,\beta))^2/\Delta^2}\right) \bigg) e^{-2(f-f_{\textrm{shift}})^2/\sigma^2}
\end{aligned}
\end{equation}
The experimental spectrum is fitted by $S(\alpha,\beta,f_{\textrm{shift}})$ with least square  to find the optimized azimuth angles $(\alpha,\beta)$.

\end{methods}

\bibliography{ref}

\bibliographystyle{naturemag}

\begin{addendum}

\item [Acknowledgements]

The authors thank J. Wrachtrup's many helpful advises on improving the work and manuscript. We also thank F. Jelezko, Y.F. Yao and F. Reinhard for their helpful discussions. This work is supported by the 973 Program
(Grant No. 2013CB921800, No. 2016YFA0502400),
the NNSFC (Grants No. 11227901, No. 31470835, and No. 91636217), the CAS (Grant No. XDB01030400, No. QYZDYSSW-SLH004, 2015370), CEBioM and the Fundamental Research
Funds for the Central Universities (WK2340000064)

\item[Author contributions]

J.D. supervised the entire project. J.D. and F.S. designed the experiments. X.K., Z.Y., P.W., and F.S. prepared the setup. N.R. and J.M. prepared the NV centers by ion implantation. X.K. and Z.Y. performed the experiments. X.K. and F.S. performed the simulation. J.D., F.S., and X.K. wrote the manuscript. All authors discussed the results and commented on the manuscript.

\item[Competing Interests] The authors declare that they have no competing financial interests.
\item[Additional information]
 Supplementary information accompanies the paper on http://www.nature.com/nat.
Correspondence and requests for materials should be
addressed to J.D.
\end{addendum}

\clearpage

\end{document}